\documentclass[twocolumn,notitlepage,superscriptaddress,showkeys,showpacs]{revtex4-1}
\usepackage{graphicx}
\usepackage{dcolumn}
\usepackage{booktabs,bm,color,braket}
\usepackage{amsmath}
\usepackage{color}

\usepackage[hidelinks,colorlinks,linkcolor=blue,citecolor=blue,urlcolor=blue]{hyperref}

\date{\today}

\begin{document}

\title{Detection of the Orbital Hall Effect by the Orbital-Spin Conversion}
\author{Jiewen Xiao}
\author{Yizhou Liu}
\author{Binghai Yan}\email{binghai.yan@weizmann.ac.il}
\affiliation{Department of Condensed Matter Physics, Weizmann Institute of Science, Rehovot 7610001, Israel}
\begin{abstract}
The intrinsic orbital Hall effect (OHE), the orbital counterpart of the spin Hall effect, was predicted and studied theoretically for more than one decade, yet to be observed in experiments. Here we propose a strategy to convert the orbital current in OHE to the spin current via the spin-orbit coupling from the contact. Furthermore, we find that OHE can induce large nonreciprocal magnetoresistance when employing the magnetic contact. Both the generated spin current and the orbital Hall magnetoresistance can be applied to probe the OHE in experiments and design orbitronic devices.
\end{abstract}
\maketitle

\section{Introduction}
The intrinsic orbital Hall effect (OHE), where an electric field induces a transverse orbital current, was proposed by the Zhang group \cite{Bernevig2005} soon after the prediction of the intrinsic spin Hall effect (SHE) \cite{murakami2003dissipationless,sinova2004universal}. The SHE was soon observed \cite{kato2004observation,wunderlich2005experimental}, later applied for the spintronic devices\cite[and references therein]{jungwirth2012spin}, and also led to the seminal discovery of the quantum SHE, i.e., the 2D topological insulator \cite{kane2005quantum,bernevig2006quantum}. Different from the SHE, the OHE does not rely on the spin-orbit coupling (SOC), and thus, it was predicted to exist in many materials \cite{Bernevig2005,Guo2005,Kontani2008a,Kontani2008b,Tanaka2008,Tokatly2010,Go2018,Jo2018,Phong2019,canonico2020orbital,Bhowal2020} with either weak or strong SOC, for example, in metals Al, Cu, Au, and Pt. 

In an OHE device, the transverse orbital current leads to the orbital accumulation at transverse edges, similar to the spin accumulation in a SHE device. Zhang \textit{et al} \cite{Bernevig2005} proposed to measure the edge orbital accumulation by the Kerr effect. Recently, Ref.~\onlinecite{go2020orbital} predicted the orbital torque generated by the orbital current. However, the OHE is yet to be detected in experiments until today. The detection of the orbital is rather challenging, because the orbital is highly non-conserved compared to the spin, especially at the device boundary.  

 A very recent work by us proposed \cite{Liu2020chiral} that the longitudinal current through DNA-type chiral materials is orbital-polarized, and contacting DNA to a large-SOC material can transform the orbital current into the spin current. Thus, we are inspired to conceive a similar way to detect the transverse OHE by converting the orbital to the spin by the SOC proximity.

In this article, we propose two ways to probe the OHE, where the strong SOC from the contact transforms the orbitronic problem to the spintronic measurement. One way is to generate spin current or spin polarization from the transverse orbital current by connecting the edge to a third lead with the strong interfacial SOC. Then the edge spin polarization and spin current is promising to be measured by the Kerr effect~\cite{kato2004observation} and the inverse SHE~\cite{saitoh2006conversion,valenzuela2006direct,Zhao2006}, respectively. The other way is to introduce a third magnetic lead and measure the magnetoresistance. We call it the orbital Hall magnetoresistance (OHMR), similar to the spin Hall magnetoresistance~\cite{Huang2012,Weiler2012}. In our proposal, the OHE refers to orbitals that resemble atomic-like orbitals, which naturally couple to the spin via the atomic SOC. We first demonstrate detection principles in a lattice model by transport calculations. Then we incorporate these principles into the metal copper, which has negligible SOC and avoids the co-existence of the SHE, as a typical example of realistic materials. In the copper based device, we demonstrate the resultant spin polarization/current and very large OHMR ($0.3 \sim 1.3 ~\%$), which are measurable by present experiment techniques.

\begin{figure}
    \centering
    \includegraphics[width=\linewidth]{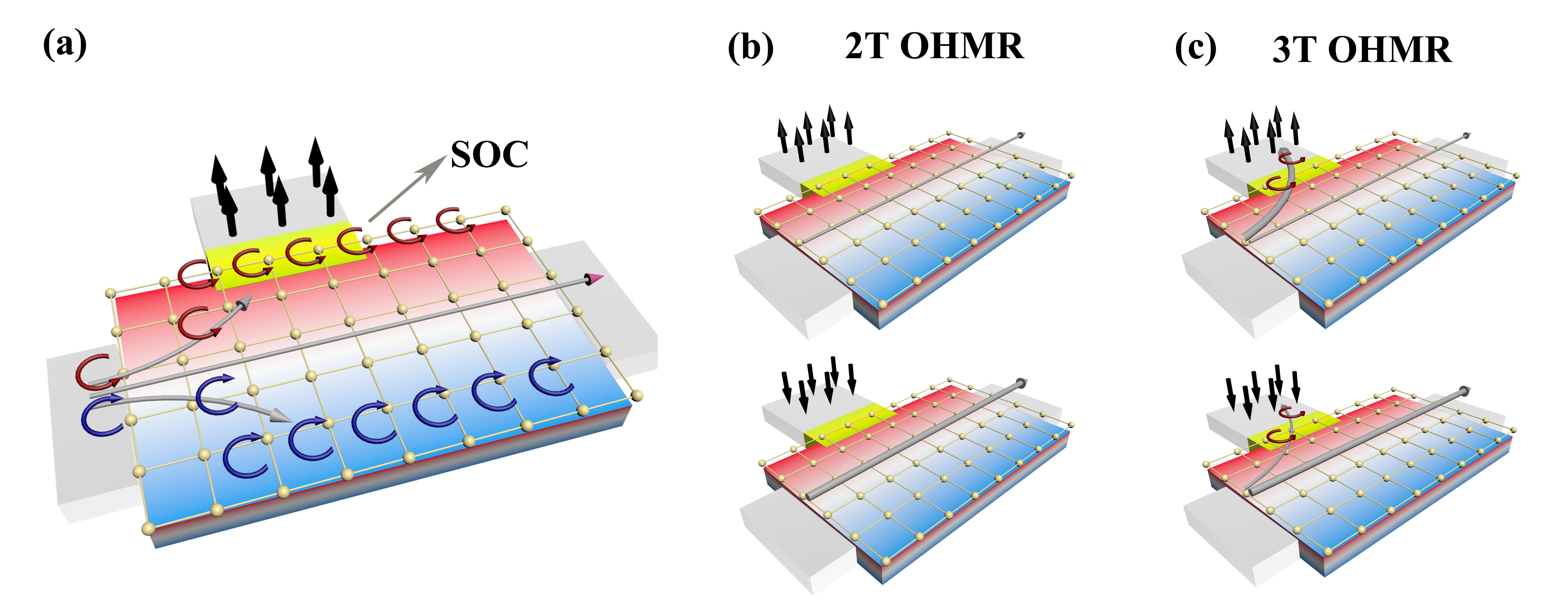}
    \caption{\textbf{Illustration of the orbital-spin conversion and the orbital Hall magnetoresistance (OHME).} 
    (a) The orbital Hall effect and the spin polarization/current generation. Opposite orbitals (red and blue circular arrows) from the left lead deflect into opposite boundaries. The red and blue backgrounds represent the orbital accumulation at two sides. Because of the SOC region (yellow) at one side, the orbital current is converted into the spin current (indicated by black arrows).
    (b) The two-terminal (2T) OHMR. The third lead is magnetized but open. (c) The three-terminal (3T) OHMR. The third lead is magnetized and conducts current. The 2T/3T conductance between different leads relies on the magnetization sensitively.  The thickness of grey curves represent the relative magnitude of the conductance.}
    \label{figure1}
\end{figure}

\section{Results and Discussions}
\subsection{Methods and General Scenario}
To detect the OHE, we introduce an extra contact with the strong SOC on the boundary of the OHE material, as shown in Figure 1. This device can act for both two-terminal (2T) and three-terminal (3T) measurements (or more terminals). 
In theoretical calculations, we completely exclude SOC from all leads so that we can well define the spin current.
We also remove SOC in the OHE material, the device regime in the center, to avoid the existence of SHE. 
Only finite atomic SOC is placed in the interfacial region (highlighted by yellow in Figure 1) between the OHE and the third lead.

We first prove the principle by a simple square-lattice model that hosts OHE. As shown in the inset of Figure 2(a), a tight-binding spinless model is constructed, with three orbitals $s$, $p_x$ and $p_y$ assigned to each site.
Under the above basis, the atomic orbital angular momentum operator $\hat{L}_z$ is written as 
\begin{equation}
\hat{L}_z = \hbar
\begin{bmatrix}
0 & 0 & 0\\
0 & 0 & -i \\
0 & i & 0 
\end{bmatrix}
\end{equation}
And three eigenstates $p_{\pm}\equiv(p_x \pm i p_y)/\sqrt{2}, s$ correspond to eigenvalues $L_z = \pm1, 0$, respectively.
After considering the nearest neighboring hopping, the Hamiltonian is written as
\begin{widetext}
\begin{equation}
H(k_x,k_y)=
\left(
\begin{array}{ccc}
E_s+2t_s\cos{k_xa}+2t_s\cos{k_ya} & -2it_{sp}\sin{k_xa} & -2it_{sp}\sin{k_ya}\\
2it_{sp}\sin{k_xa} & E_{px}+2t_{p\sigma}\cos{k_xa}+2t_{p\pi}\cos{k_ya} & 0\\
2it_{sp}\sin{k_ya} & 0 & E_{py}+2t_{p\pi}\cos{k_xa}+2t_{p\sigma}\cos{k_ya}\\
\end{array}
\right) 
\end{equation}
\end{widetext}
where $E_s$, $E_{px}$ and $E_{py}$ are onsite energies of $s$, $p_x$ and $p_y$ orbitals. $t_s$, $t_{p\sigma}$, $t_{p\pi}$, $t_{sp}$ are electron hopping integrals between $s$ orbitals, $\sigma$ type oriented $p$ orbitals, $\pi$ type oriented $p$ orbitals, and $s$ and $p$ orbitals, respectively. In the following calculations, their values are specified as $E_s=1.3$, $E_{px}=E_{py}=-1.9$, $t_s=-0.3$, $t_{p\sigma}=0.6$, $t_{p\pi}=0.3$, and $t_{sp}=0.5$, in the unit of eV.
To realize the OHE, it requires the inter-orbital hopping to induce the transverse $L_z$ current.
Since $p_x$ and $p_y$ orbital are orthogonal under the square lattice geometry, the inter-orbital hopping $t_{sp}$ becomes the critical parameter that controls the existence of the OHE.
Then we introduce the atomic SOC on the boundary to demonstrate the OHE detection by $\lambda_{soc}\hat{S}_z\cdot \hat{L}_z$, where $\hat{S}_z$ is the spin operator.

We estimate the OHE conductivity ($\sigma_{OH}$) with the orbital Berry curvature in the Kubo formula \cite{Xiao2010,Nagaosa2010},
\begin{equation}
    \sigma_{OH} = \frac{e}{\hbar}\sum_{n} \int \frac{d^3k}{(2\pi)^3}f_{nk}\Omega_n^{L_z}(k)
\end{equation}
\begin{equation}
    \Omega_{n}^{L_z}(k) = 2\hbar^2\sum_{m\neq n} Im[\frac{\bra{u_{nk}}j_y^{L_z}\ket{u_{mk}}\bra{u_{mk}}\hat{v}_x\ket{u_{nk}}}{(E_{nk}-E_{mk})^2}]
\end{equation}
where $\Omega_{n}^{L_z}(k)$ is the ``orbital'' Berry curvature for the $n^{th}$ band with Bloch state $\ket{u_{nk}}$ and energy eigenvalue $E_{nk}$. $f_{nk}$ is the Fermi-Dirac distribution function. $v_x$ is the $x$ component of the band velocity operator while $j_y^{L_z}$ is the orbital current operator in the $y$ direction, defined as $j_y^{L_z} = (\hat{L}_z \hat{v}_y+\hat{v}_y \hat{L}_z)/2 $. 
Therefore, the above formula indicates that the interband perturbation induces the orbital Berry curvature, further reiterating the importance of inter-orbital hopping. 
We also note that, the orbital Berry curvature is even under the time-reversal symmetry or the spatial inversion symmetry, $\Omega_{n}^{L_z}(k) = \Omega_{n}^{L_z}(-k)$.

For the device schematically presented in Figure 1, we calculated the conductance by the Landauer-B\"uttiker formula \cite{Buttiker1986} with the scattering matrix from lead $i$ to lead $j$,
\begin{equation}
    G^{i \rightarrow j} = \frac{e^2}{h}\sum\limits_{n \in j, m \in i} |S_{nm}|^2,
\end{equation}
where $S_{mn}$ is the scattering matrix element from the $m$-th eigenstate in lead $i$ to the $n^{th}$ eigenstate in lead $j$. In all three leads ($i,j=1,2,3$), spin ($S_z=\uparrow \downarrow$) is a conserved quantity because of the lack of SOC. 
We turn off the inter-orbital hopping in leads so that $L_z$ is also conserved, i.e., $L_z$ commutes with the Hamiltonian (See Supplementary Materials). Therefore, with the spin and orbital conserved leads, we can specify the conductance in each $S_z$ and $L_z$ channel, and define the orbital- and spin-polarized conductance as:
\begin{align}
G^{ij}_{S_z} &= G^{i \rightarrow j\uparrow}-G^{j \rightarrow j\downarrow} \\
G^{ij}_{L_z} &= G^{i \rightarrow j+}-G^{i \rightarrow j-},
\end{align}
where $G^{ij}_{S_z (L_z)}$ is the conductance from lead $i$ to the $S_z (L_z)$ channel of lead $j$. 
$G^{i \rightarrow j0}$ is omitted here since $L_z = 0$ contributes no polarization. We performed the conductance calculations with the quantum transport package Kwant \cite{Groth2014kwant}.

As illustrated in Figure 1(a), electrons with the opposite orbital angular momentum deflect into transverse directions in the OHE region, resulting in the transverse orbital current. Therefore, orbital accumulates at two sides, and the orbital polarization emerges. 
To detect the orbital polarization, atomic SOC is added at one side, as highlighted by yellow in Figure 1(a).
After electron deflecting into the SOC region, the right-handed orbital (red circular arrows) is converted to the up spin polarization.
If a third lead is further attached, the SOC region converts the orbital current into the spin current. If the third lead exhibits magnetization along $z$ ($M_z$) (Figure 1(b) and 1(c)), inversely, the OHE induces the OHMR, relying on whether $M_z$ is parallel or anti-parallel to the generated spin polarization. 
In the 2T measurement (Figure 1(b)), the conductance from lead 1 to lead 2 ($G_{1\rightarrow2}$) changes when the $M_z$ direction is reversed. And the changing direction of $G_{1\rightarrow2}$ depends sensitively on the size of the device, due to the complex orbital accumulation and reflection with an open lead. While for its spin counterpart, the SHE-induced magnetoresistance is commonly measured in a 2T setup~\cite{Huang2012,Weiler2012}. 
In the 3T device (Figure 1(c)), the situation is simpler since the transverse orbital current can flow into the third lead. If $M_z$ and spin polarization is parallel (anti-parallel), the transverse orbital current matches (mismatches) the lead magnetization, resulting in the high (low) $G_{1\rightarrow 3}$ and low (high) $G_{1\rightarrow2}$ accordingly. 
We point out that the 3T measurement is usually more favorable than 2T, since the 3T device avoids the 2T reciprocity constrain \cite{Buttiker1988} and the conductance change [$\Delta G = G(M_z) - G(-M_z)$] is also relatively larger in the third lead, as discussed in the following.

\subsection{Spin Polarization and Spin Current Generated by the OHE}

The band structure weighted by the orbital Berry curvature for the square lattice is plotted in Figure 2(a). 
The highest band corresponds to the $s$ orbital dispersion, while two lower bands are dominated by $p$ orbitals. 
The orbital Berry Curvature concentrates near the $\Gamma$ point, $M$ point and $\Gamma$-$M$ line in the Brillouin zone, where band hybridization is strong. 
After integrating $\Omega_{Lz}$ in the Brillouin zone, the orbital Hall conductivity is derived and presented in Figure 2(a). 
It shows that, due to the inter-orbital hopping $t_{sp}$, states below ($p$ orbitals) and above ($s$ orbital) Fermi level both exhibits significant $\sigma_{OH}$. 
However, if $t_{sp}$ is turned off so that $L_z$ is conserved, both $\Omega_{Lz}$ and $\sigma_{OH}$ vanishes. 

Based on the square lattice with finite $t_{sp}$, the 2T device is constructed, as shown in Figure 2(b). 
Without SOC at two sides, the orbital density distribution is plotted in Figure 2(c), which shows that opposite orbitals accumulate and polarize at two boundaries.
With SOC turned on, spin density appears and largely concentrates on the local SOC atoms, which is promising to be detected by the Kerr effect \cite{kato2004observation}.
Since SOC couples the $p_{+}$ ($p_-$) orbital to the $\uparrow$ ($\downarrow$) spin and forms the $\ket{j_m = \frac{3}{2}}$ ($\ket{j_m = -\frac{3}{2}}$) state, the spin density near the SOC region largely follows the orbital density pattern: positive at the upper side and negative at the lower side. 
To verify that the spin polarization is directly induced by the OHE rather than SOC, we turned off the OHE by setting $t_{sp}=0$ eV and preserve the SOC at the interface. The supplementary Figure S2 shows that both the orbital and spin polarization disappear.

On the basis of 2T device, a third lead is attached to the SOC side to form a 3T device, as shown in Figure 2(b). Therefore, rather than the orbital accumulation, the orbital current will flow into the third lead and generate the spin current.
Figure 2(d) shows that the orbital current from lead 1 to lead 3 ($G^{13}_{L_z}$) exists with and without SOC at the interface.  
For instance, for states above Fermi level, $L_z = +1$ states are more easily transported into lead 3 than $L_z = -1$ states, and thus polarizes the lead, being consistent with the positive orbital polarization at the upper side in Figure 2(c).
On the other hand, for the spin conductance, it only appears when turning on SOC, and the energy dependence of $G^{13}_{S_z}$ largely follows the orbital conductance, further demonstrating the spin generation process from the orbital.  
If we increase the SOC strength, $G^{13}_{S_z}$ increases accordingly, because of the higher orbital-spin conversion efficiency (see Figure S3). 
We also test orbital non-conserved leads with nonzero $t_{sp}$, whose spin conductance remains the similar feature (see Figure S4).
\begin{figure}
    \centering
    \includegraphics[width=\linewidth]{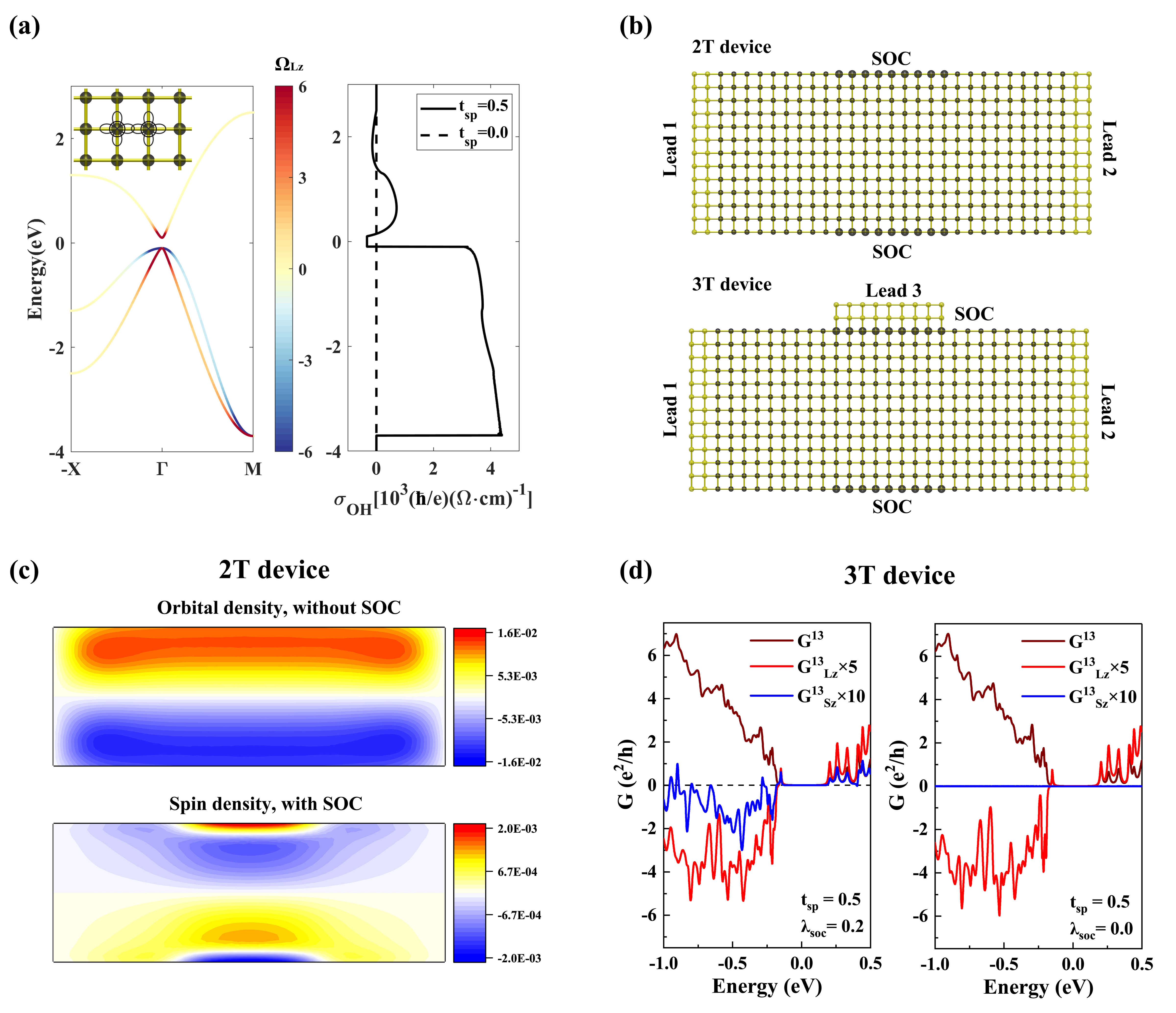}
    \caption{\textbf{Orbital-spin conversion in the two terminal (2T) and three terminal (3T) device.} (a) Band structure of the square lattice with $t_{sp}=0.5$ eV (left) and the orbital Hall conductivity with $t_{sp}=0.5$ eV and $t_{sp}=0.0$ eV (right). In the inset, the tight binding model of the square lattice is presented. (b) 2T and 3T detection devices, where larger spheres at two sides represent SOC regions. The yellow spheres at left, right and upper sides represent leads. (c) Orbital and spin density distribution in the 2T setup, at the energy level of 0.2 eV. (d) Total, orbital and spin conductance from lead 1 to lead 3 with (left) and without (right) SOC.}
    \label{figure2}
\end{figure}

\begin{figure*}
    \centering
    \includegraphics[width=14cm]{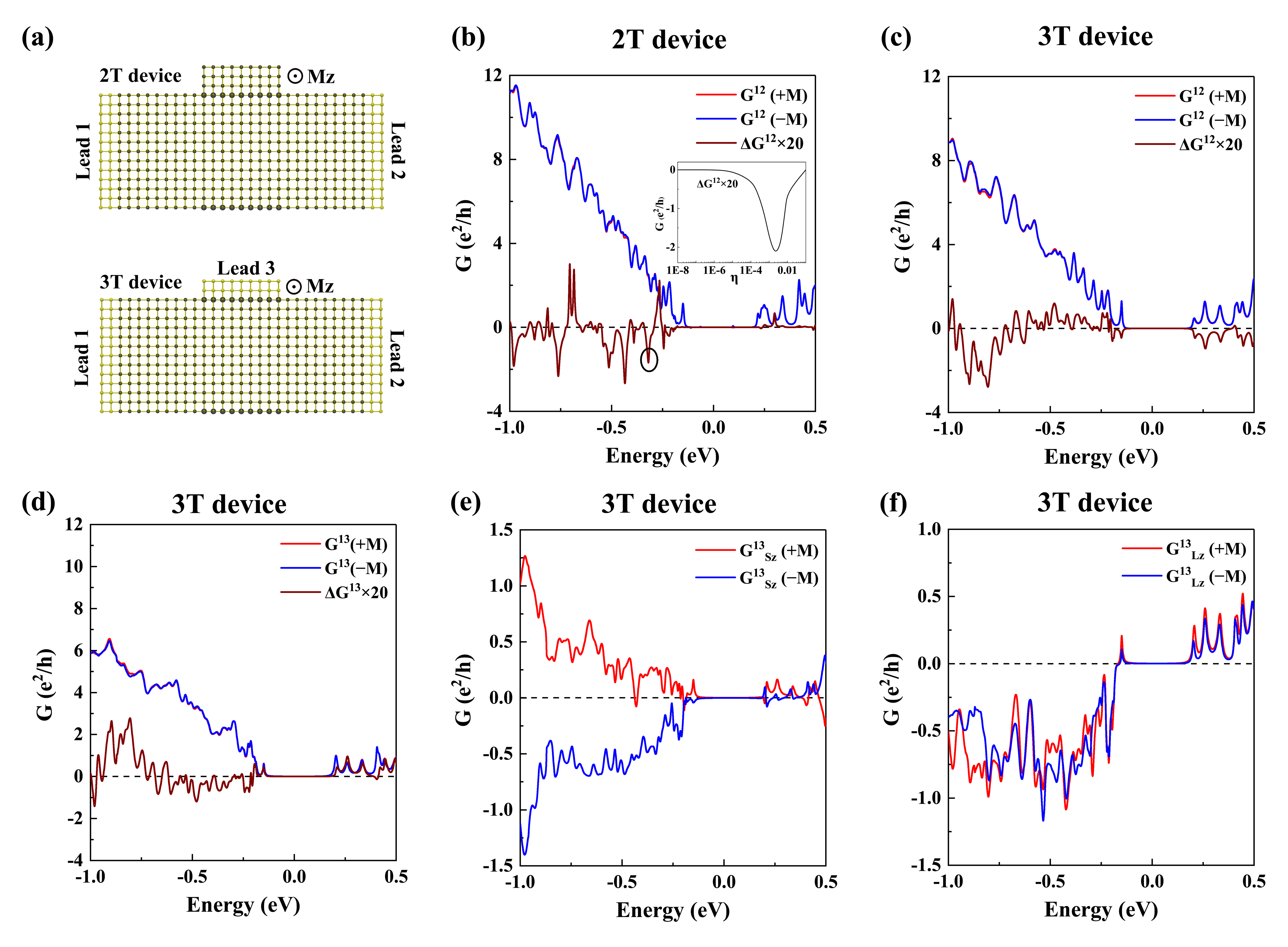}
    \caption{\textbf{Orbital Magnetoresistance with magnetic leads.} 
    (a) 2T and 3T detection devices with the exchange field $\pm M_z$ in the open and conducting lead 3. Larger spheres represent the interfacial SOC region. (b) Total conductance from lead 1 to lead 2 in $\pm M_z$ field in the 2T device, with the dephasing term $\eta$ set to 0.001. In the inset, the dephasing dependent $\Delta G_{12}$ for the peak inside the circle is presented. (c) Total conductance from lead 1 to lead 2 in $\pm M_z$ field in the 3T device. (d) Total, (e) Spin and (f) Orbital conductance from lead 1 to lead 3 in $\pm M_z$ field in the 3T device. In all these calculations, $\lambda_{SOC}$ and $M_z$ is set to 0.2 eV and 0.4 eV, respectively.
    }
    \label{figure3}
\end{figure*}

\subsection{Orbital Hall Magnetoresistance}

As discussed above, the current injected into lead 3 is spin-polarized. When lead 3 is magnetized along the $z$ axis, we expect the existence of magnetization-dependent conductance, i.e. $G^{13}(M_z) \neq G^{13}(-M_z)$. From the current conservation \cite{Buttiker1988}, we deduce the relation,
\begin{align}
\Delta G^{13} &= -\Delta G^{12}
\end{align}
where $\Delta G^{ij} \equiv G^{ij}(M_z)-G^{ij}(-M_z)$. 
To demonstrate this, $M_z$ is introduced to lead 3 as an exchange field to the spin, as shown in the 3T setup in Figure 3(a). 
Results in Figure 3(c) and 3(d) indicate that $\Delta G^{12}$ and $\Delta G^{13}$ can reach several percentage of the total conductance at some energies.  
We also confirm that $\Delta G^{12}$ and $\Delta G^{13}$ are proportional to the exchange field strength (see Figure S5).

To understand the orbital induced magnetoresistance, the spin and orbital conductance from lead 1 to lead 3 are calculated. 
As shown in Figure 3(e), $G^{13}_{S_z}$ almost changes its sign when flipping $M_z$ in lead 3, as expected. And $G^{13}_{S_z}$ now inversely affects $G^{13}_{L_z}$ because of the interfacial SOC. When further comparing Figure 3(d) and 3(f), we found that the change of the magnitude of $G^{13}_{L_z}$ is proportional to the change of total conductance $\Delta G^{13}$. Therefore, it verifies the scenario in Figure 1(c): when the orbital matches the spin in magnetic leads, $G^{13}_{L_z}$ and thus $G^{13}$ is higher while $G^{12}$ is accordingly lower. Thus, it indicates the essential role of the orbital in connecting charge and spin in the transport.

However, the 2T results exhibit qualitatively different features from the 3T results. According to the reciprocity relation \cite{Buttiker1988}, the 2T conductance obeys $G^{12}(M_z) = G^{12}(-M_z)$. Only when the current conservation is broken, one may obtain the 2T magentoresistance. Therefore, we introduce a dephasing term $i \eta$ to leak electrons into virtual leads \cite{Buttiker1986b} to release the above constrain. As shown in the inset of Figure 3(b), the $\Delta G^{12}$ is zero at $\eta =0$, first increases quickly and soon decreases as further increasing $\eta$. In the large $\eta$ limit, the system is totally out of coherence and thus, the conductance cannot remember the spin and orbital information. We note that the dephasing exists ubiquitously in experiments due to the dissipative scattering for example by electron-phonon interaction and impurities. 

For the same $M_z$, the 2T $\Delta G^{12}$ (Figure (3b)) roughly exhibits the opposite sign compared to the 3T $\Delta G^{12}$ (Figure (3c)) in the energy window investigated. Unlike that $\Delta G^{13}$ follows the change of $G^{13}_{L_z}$ (Figure (3f)), the change direction of $G_{12}$ depends on the geometry of the 2T device (see Figure S6). The magnitude of the 2T $\Delta G^{12}$ also depends sensitively on the value of $\eta$. Its peak value (Figure (3b)), with $\eta$ around 0.001, is comparable with the 3T value in the same parameter regime. However, the 3T conductance avoids the strict constrain of the 2T reciprocity, and the existence of 3T OHMR does not rely on the dephasing. Furthermore, in the 3T setup, the magnetoresistance ratio $\Delta G_{13}/G_{13}$ in the third lead is also larger than $\Delta G_{12}/G_{12}$, because of the lower total conductance of $G_{13}$. Therefore, we propose that the 3T setup may be more advantageous to detect the OHE.

\subsection{Realistic Material Cu}

\begin{figure*}
    \centering
    \includegraphics[width=14cm]{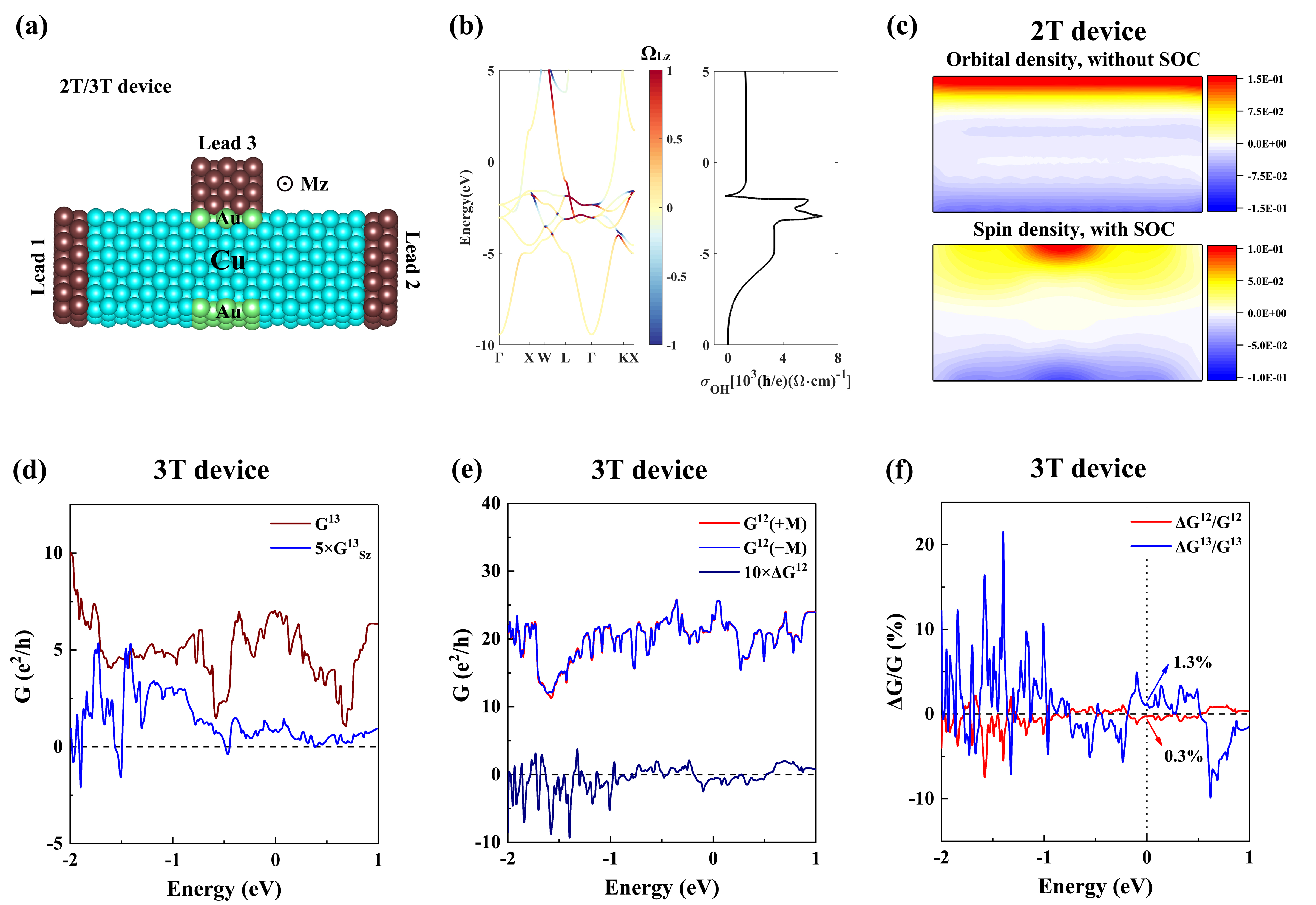}
    \caption{\textbf{The orbital-spin conversion and orbital magnetoresistance in Cu.}
    (a) 2T/3T device based on real materials, where the scattering region and leads are treated as Cu and the SOC region is treated as Au. (b) The band structure of Cu, weighted by the orbital Berry curvature $\Omega_{Lz}$, and the corresponding orbital Hall conductivity. (c) Orbital and spin density distribution at the Fermi level in the 2T setup. (d) Spin conductance from lead 1 to lead 3 in nonmagnetic leads in the 3T setup. (e) Total conductance from lead 1 to lead 2 in $\pm M_z$ exchange field in the 3T setup. (f) Magnetoresistance ratio $\Delta G^{12} / G^{12}$ and $\Delta G^{13} / G^{13}$ in the 3T setup.}
    \label{figure4}
\end{figure*}

Based on the simple square lattice model, we demonstrate two main phenomena, the OHE-induced spin polarization / spin current current assisted by the atomic SOC on the boundary and the existence of OHMR. 
We further examine them in a realistic material Cu. This light noble metal is predicted to exhibit the strong OHE.

As shown in Figure 4(a), the 2T (without lead 3) and 3T devices are composed of Cu (without SOC) in both the scattering region and leads, and the heavy metal Au (with SOC) at two boundaries. We adopted the tight-binding method to describe the Cu and leads, where 9 atomic orbitals ($s$, $p_x$, $p_y$, $p_z$, $d_{z^2}$, $d_{x^2-y^2}$, $d_{xy}$, $d_{yz}$, $d_{zx}$) are assigned to each site. The nearest-neighboring and the second-nearest-neighboring hoppings are considered with the Slater-Koster type parameters from Ref.~\onlinecite{handbook1986}. For the heavy metal Au at two sides, the SOC strength is set to 0.37 eV as suggested by Ref.~\onlinecite{handbook1986}. With the tight-binding approach, the first-principles band structure of Cu is reproduced (see Figure S7). 

As shown in Figure 4(b), the orbital Berry curvature concentrates on the $d$ orbital region ($-4$ eV $\sim -2$ eV) due to the orbital hybridization, consistent with previous works~\cite{Tanaka2008,Jo2018}. 
After integrating $\Omega_{Lz}$, Figure 4(b) shows that the orbital Hall conductivity is around 6000 ${(\hbar / e)(\Omega cm)}^{-1}$ in the $d$ orbital region, even larger than the spin Hall conductivity of Pt. 
Near the Fermi level, the orbital Hall conductivity is determined by the $s$-orbital derived bands and reduces to around 1000 ${(\hbar / e)(\Omega cm)}^{-1}$. 

For the 2T device, the orbital and spin density at Fermi level are plotted in Figure 4(c). The orbital polarization exists at two sides as a consequence of the OHE. With the heavy metal Au attached, spin polarization is generated, which concentrates on Au atoms and follows the orbital density pattern. 
To confirm that the spin polarization is induced by the OHE, we artificially turn off the inter-orbital hopping in Cu to eliminate the OHE, but still keep the SOC in the Au region. Result show that both the orbital and spin polarization disappear (see Figure S8), in accordance with our prediction.

For the 3T device, we add a third Cu lead to one SOC side and calculate the spin conductance from lead 1 to lead 3 ($G^{13}_{S_z}$). 
As shown in Figure 4(d), the generated spin conductance displays an energy-dependence similar to the bulk $\sigma_{OH}$. 
Near the Fermi level, the spin polarization rate can reach 4 $\%$, and it is even around 20$\%$ in the $d$ orbital region. Therefore, a sizable spin current can also be generated from the OHE by adding an interfacial SOC layer. Similarly, when artificially switching off the OHE of Cu but keeping the Au part, the spin current disappears, eliminating the contribution of the SHE brought by the thin Au layer (see Figure S9). 

We also studied the OHMR by applying an exchange field $M_z$ in the lead 3. We choose $M_z = 0.95$ eV according to the approximate spin splitting in the transition metal Co (see Figure S10).  
As shown in Figure 4(e) and 4(f), the 3T OHMR is rather large, where we find $\Delta G^{12} / G^{12} \approx 0.3\% $ and $\Delta G^{13} / G^{13} \approx 1.3 \% $ at the Fermi level. 
In experiment, the SHE magentoresistance is around $0.05\sim 0.5 \%$ (see Ref.~\onlinecite{Avci2015} for example). Therefore, the sizable OHMR in copper can be fairly measurable by present experimental techniques. 
We should point out that similar effects can be generalized to other OHE
materials like Li and Al \cite{Jo2018}.

\section{Summary}
In summary, we have proposed the OHE detection strategies by converting the orbital to spin by the interfacial SOC, and inducing the strong spin current/polarization. Inversely, the OHE can also generate the large nonreciprocal magnetoresistance when employing the magnetic contact. 
We point out that, compared to the two-terminal one, the three-terminal OHMR does not require the dephasing term , and may be more advantageous to detect the OHE. 
Using the device setup based on the metal Cu, we demonstrate that the generated spin polarization and OHMR are strong enough to be measured in the present experimental condition. 
Our work will pave a way to realize the OHE in experiment, and further design orbitronic or even orbitothermal devices for future applications.

\section{Acknowledgement}
We honor the memory of Prof. Shoucheng Zhang. This article follows his earlier works on the intrinsic orbital Hall effect and spin Hall effect. B.Y. acknowledges the financial support by the Willner Family Leadership Institute for the Weizmann Institute of Science, the Benoziyo Endowment Fund for the Advancement of Science,  Ruth and Herman Albert Scholars Program for New Scientists, and the European Research Council (ERC) under the European Union's Horizon 2020 research and innovation programme (Grant No. 815869, NonlinearTopo).


\begin{thebibliography}{33}%
\makeatletter
\providecommand \@ifxundefined [1]{%
 \@ifx{#1\undefined}
}%
\providecommand \@ifnum [1]{%
 \ifnum #1\expandafter \@firstoftwo
 \else \expandafter \@secondoftwo
 \fi
}%
\providecommand \@ifx [1]{%
 \ifx #1\expandafter \@firstoftwo
 \else \expandafter \@secondoftwo
 \fi
}%
\providecommand \natexlab [1]{#1}%
\providecommand \enquote  [1]{``#1''}%
\providecommand \bibnamefont  [1]{#1}%
\providecommand \bibfnamefont [1]{#1}%
\providecommand \citenamefont [1]{#1}%
\providecommand \href@noop [0]{\@secondoftwo}%
\providecommand \href [0]{\begingroup \@sanitize@url \@href}%
\providecommand \@href[1]{\@@startlink{#1}\@@href}%
\providecommand \@@href[1]{\endgroup#1\@@endlink}%
\providecommand \@sanitize@url [0]{\catcode `\\12\catcode `\$12\catcode
  `\&12\catcode `\#12\catcode `\^12\catcode `\_12\catcode `\%12\relax}%
\providecommand \@@startlink[1]{}%
\providecommand \@@endlink[0]{}%
\providecommand \url  [0]{\begingroup\@sanitize@url \@url }%
\providecommand \@url [1]{\endgroup\@href {#1}{\urlprefix }}%
\providecommand \urlprefix  [0]{URL }%
\providecommand \Eprint [0]{\href }%
\providecommand \doibase [0]{http://dx.doi.org/}%
\providecommand \selectlanguage [0]{\@gobble}%
\providecommand \bibinfo  [0]{\@secondoftwo}%
\providecommand \bibfield  [0]{\@secondoftwo}%
\providecommand \translation [1]{[#1]}%
\providecommand \BibitemOpen [0]{}%
\providecommand \bibitemStop [0]{}%
\providecommand \bibitemNoStop [0]{.\EOS\space}%
\providecommand \EOS [0]{\spacefactor3000\relax}%
\providecommand \BibitemShut  [1]{\csname bibitem#1\endcsname}%
\let\auto@bib@innerbib\@empty
\bibitem [{\citenamefont {Bernevig}\ \emph {et~al.}(2005)\citenamefont
  {Bernevig}, \citenamefont {Hughes},\ and\ \citenamefont
  {Zhang}}]{Bernevig2005}%
  \BibitemOpen
  \bibfield  {author} {\bibinfo {author} {\bibfnamefont {B.~A.}\ \bibnamefont
  {Bernevig}}, \bibinfo {author} {\bibfnamefont {T.~L.}\ \bibnamefont
  {Hughes}}, \ and\ \bibinfo {author} {\bibfnamefont {S.-C.}\ \bibnamefont
  {Zhang}},\ }\href {\doibase 10.1103/physrevlett.95.066601} {\bibfield
  {journal} {\bibinfo  {journal} {Physical Review Letters}\ }\textbf {\bibinfo
  {volume} {95}},\ \bibinfo {pages} {066601} (\bibinfo {year}
  {2005})}\BibitemShut {NoStop}%
\bibitem [{\citenamefont {Murakami}\ \emph {et~al.}(2003)\citenamefont
  {Murakami}, \citenamefont {Nagaosa},\ and\ \citenamefont
  {Zhang}}]{murakami2003dissipationless}%
  \BibitemOpen
  \bibfield  {author} {\bibinfo {author} {\bibfnamefont {S.}~\bibnamefont
  {Murakami}}, \bibinfo {author} {\bibfnamefont {N.}~\bibnamefont {Nagaosa}}, \
  and\ \bibinfo {author} {\bibfnamefont {S.-C.}\ \bibnamefont {Zhang}},\
  }\href@noop {} {\bibfield  {journal} {\bibinfo  {journal} {Science}\ }\textbf
  {\bibinfo {volume} {301}},\ \bibinfo {pages} {1348} (\bibinfo {year}
  {2003})}\BibitemShut {NoStop}%
\bibitem [{\citenamefont {Sinova}\ \emph {et~al.}(2004)\citenamefont {Sinova},
  \citenamefont {Culcer}, \citenamefont {Niu}, \citenamefont {Sinitsyn},
  \citenamefont {Jungwirth},\ and\ \citenamefont
  {MacDonald}}]{sinova2004universal}%
  \BibitemOpen
  \bibfield  {author} {\bibinfo {author} {\bibfnamefont {J.}~\bibnamefont
  {Sinova}}, \bibinfo {author} {\bibfnamefont {D.}~\bibnamefont {Culcer}},
  \bibinfo {author} {\bibfnamefont {Q.}~\bibnamefont {Niu}}, \bibinfo {author}
  {\bibfnamefont {N.}~\bibnamefont {Sinitsyn}}, \bibinfo {author}
  {\bibfnamefont {T.}~\bibnamefont {Jungwirth}}, \ and\ \bibinfo {author}
  {\bibfnamefont {A.~H.}\ \bibnamefont {MacDonald}},\ }\href@noop {} {\bibfield
   {journal} {\bibinfo  {journal} {Physical review letters}\ }\textbf {\bibinfo
  {volume} {92}},\ \bibinfo {pages} {126603} (\bibinfo {year}
  {2004})}\BibitemShut {NoStop}%
\bibitem [{\citenamefont {Kato}\ \emph {et~al.}(2004)\citenamefont {Kato},
  \citenamefont {Myers}, \citenamefont {Gossard},\ and\ \citenamefont
  {Awschalom}}]{kato2004observation}%
  \BibitemOpen
  \bibfield  {author} {\bibinfo {author} {\bibfnamefont {Y.~K.}\ \bibnamefont
  {Kato}}, \bibinfo {author} {\bibfnamefont {R.~C.}\ \bibnamefont {Myers}},
  \bibinfo {author} {\bibfnamefont {A.~C.}\ \bibnamefont {Gossard}}, \ and\
  \bibinfo {author} {\bibfnamefont {D.~D.}\ \bibnamefont {Awschalom}},\
  }\href@noop {} {\bibfield  {journal} {\bibinfo  {journal} {science}\ }\textbf
  {\bibinfo {volume} {306}},\ \bibinfo {pages} {1910} (\bibinfo {year}
  {2004})}\BibitemShut {NoStop}%
\bibitem [{\citenamefont {Wunderlich}\ \emph {et~al.}(2005)\citenamefont
  {Wunderlich}, \citenamefont {Kaestner}, \citenamefont {Sinova},\ and\
  \citenamefont {Jungwirth}}]{wunderlich2005experimental}%
  \BibitemOpen
  \bibfield  {author} {\bibinfo {author} {\bibfnamefont {J.}~\bibnamefont
  {Wunderlich}}, \bibinfo {author} {\bibfnamefont {B.}~\bibnamefont
  {Kaestner}}, \bibinfo {author} {\bibfnamefont {J.}~\bibnamefont {Sinova}}, \
  and\ \bibinfo {author} {\bibfnamefont {T.}~\bibnamefont {Jungwirth}},\
  }\href@noop {} {\bibfield  {journal} {\bibinfo  {journal} {Physical review
  letters}\ }\textbf {\bibinfo {volume} {94}},\ \bibinfo {pages} {047204}
  (\bibinfo {year} {2005})}\BibitemShut {NoStop}%
\bibitem [{\citenamefont {Jungwirth}\ \emph {et~al.}(2012)\citenamefont
  {Jungwirth}, \citenamefont {Wunderlich},\ and\ \citenamefont
  {Olejn{\'\i}k}}]{jungwirth2012spin}%
  \BibitemOpen
  \bibfield  {author} {\bibinfo {author} {\bibfnamefont {T.}~\bibnamefont
  {Jungwirth}}, \bibinfo {author} {\bibfnamefont {J.}~\bibnamefont
  {Wunderlich}}, \ and\ \bibinfo {author} {\bibfnamefont {K.}~\bibnamefont
  {Olejn{\'\i}k}},\ }\href@noop {} {\bibfield  {journal} {\bibinfo  {journal}
  {Nature materials}\ }\textbf {\bibinfo {volume} {11}},\ \bibinfo {pages}
  {382} (\bibinfo {year} {2012})}\BibitemShut {NoStop}%
\bibitem [{\citenamefont {Kane}\ and\ \citenamefont
  {Mele}(2005)}]{kane2005quantum}%
  \BibitemOpen
  \bibfield  {author} {\bibinfo {author} {\bibfnamefont {C.~L.}\ \bibnamefont
  {Kane}}\ and\ \bibinfo {author} {\bibfnamefont {E.~J.}\ \bibnamefont
  {Mele}},\ }\href@noop {} {\bibfield  {journal} {\bibinfo  {journal} {Physical
  review letters}\ }\textbf {\bibinfo {volume} {95}},\ \bibinfo {pages}
  {226801} (\bibinfo {year} {2005})}\BibitemShut {NoStop}%
\bibitem [{\citenamefont {Bernevig}\ \emph {et~al.}(2006)\citenamefont
  {Bernevig}, \citenamefont {Hughes},\ and\ \citenamefont
  {Zhang}}]{bernevig2006quantum}%
  \BibitemOpen
  \bibfield  {author} {\bibinfo {author} {\bibfnamefont {B.~A.}\ \bibnamefont
  {Bernevig}}, \bibinfo {author} {\bibfnamefont {T.~L.}\ \bibnamefont
  {Hughes}}, \ and\ \bibinfo {author} {\bibfnamefont {S.-C.}\ \bibnamefont
  {Zhang}},\ }\href@noop {} {\bibfield  {journal} {\bibinfo  {journal}
  {science}\ }\textbf {\bibinfo {volume} {314}},\ \bibinfo {pages} {1757}
  (\bibinfo {year} {2006})}\BibitemShut {NoStop}%
\bibitem [{\citenamefont {Guo}\ \emph {et~al.}(2005)\citenamefont {Guo},
  \citenamefont {Yao},\ and\ \citenamefont {Niu}}]{Guo2005}%
  \BibitemOpen
  \bibfield  {author} {\bibinfo {author} {\bibfnamefont {G.~Y.}\ \bibnamefont
  {Guo}}, \bibinfo {author} {\bibfnamefont {Y.}~\bibnamefont {Yao}}, \ and\
  \bibinfo {author} {\bibfnamefont {Q.}~\bibnamefont {Niu}},\ }\href {\doibase
  10.1103/PhysRevLett.94.226601} {\bibfield  {journal} {\bibinfo  {journal}
  {Phys. Rev. Lett.}\ }\textbf {\bibinfo {volume} {94}},\ \bibinfo {pages}
  {226601} (\bibinfo {year} {2005})}\BibitemShut {NoStop}%
\bibitem [{\citenamefont {Kontani}\ \emph
  {et~al.}(2008{\natexlab{a}})\citenamefont {Kontani}, \citenamefont {Tanaka},
  \citenamefont {Hirashima}, \citenamefont {Yamada},\ and\ \citenamefont
  {Inoue}}]{Kontani2008a}%
  \BibitemOpen
  \bibfield  {author} {\bibinfo {author} {\bibfnamefont {H.}~\bibnamefont
  {Kontani}}, \bibinfo {author} {\bibfnamefont {T.}~\bibnamefont {Tanaka}},
  \bibinfo {author} {\bibfnamefont {D.~S.}\ \bibnamefont {Hirashima}}, \bibinfo
  {author} {\bibfnamefont {K.}~\bibnamefont {Yamada}}, \ and\ \bibinfo {author}
  {\bibfnamefont {J.}~\bibnamefont {Inoue}},\ }\href {\doibase
  10.1103/physrevlett.100.096601} {\bibfield  {journal} {\bibinfo  {journal}
  {Physical Review Letters}\ }\textbf {\bibinfo {volume} {100}},\ \bibinfo
  {pages} {096601} (\bibinfo {year} {2008}{\natexlab{a}})},\ \Eprint
  {http://arxiv.org/abs/cond-mat/0702447} {cond-mat/0702447} \BibitemShut
  {NoStop}%
\bibitem [{\citenamefont {Kontani}\ \emph
  {et~al.}(2008{\natexlab{b}})\citenamefont {Kontani}, \citenamefont {Tanaka},
  \citenamefont {Hirashima}, \citenamefont {Yamada},\ and\ \citenamefont
  {Inoue}}]{Kontani2008b}%
  \BibitemOpen
  \bibfield  {author} {\bibinfo {author} {\bibfnamefont {H.}~\bibnamefont
  {Kontani}}, \bibinfo {author} {\bibfnamefont {T.}~\bibnamefont {Tanaka}},
  \bibinfo {author} {\bibfnamefont {D.~S.}\ \bibnamefont {Hirashima}}, \bibinfo
  {author} {\bibfnamefont {K.}~\bibnamefont {Yamada}}, \ and\ \bibinfo {author}
  {\bibfnamefont {J.}~\bibnamefont {Inoue}},\ }\href {\doibase
  10.1103/physrevlett.102.016601} {\bibfield  {journal} {\bibinfo  {journal}
  {Physical Review Letters}\ }\textbf {\bibinfo {volume} {102}},\ \bibinfo
  {pages} {016601} (\bibinfo {year} {2008}{\natexlab{b}})},\ \Eprint
  {http://arxiv.org/abs/0806.0210} {0806.0210} \BibitemShut {NoStop}%
\bibitem [{\citenamefont {Tanaka}\ \emph {et~al.}(2008)\citenamefont {Tanaka},
  \citenamefont {Kontani}, \citenamefont {Naito}, \citenamefont {Naito},
  \citenamefont {Hirashima}, \citenamefont {Yamada},\ and\ \citenamefont
  {Inoue}}]{Tanaka2008}%
  \BibitemOpen
  \bibfield  {author} {\bibinfo {author} {\bibfnamefont {T.}~\bibnamefont
  {Tanaka}}, \bibinfo {author} {\bibfnamefont {H.}~\bibnamefont {Kontani}},
  \bibinfo {author} {\bibfnamefont {M.}~\bibnamefont {Naito}}, \bibinfo
  {author} {\bibfnamefont {T.}~\bibnamefont {Naito}}, \bibinfo {author}
  {\bibfnamefont {D.~S.}\ \bibnamefont {Hirashima}}, \bibinfo {author}
  {\bibfnamefont {K.}~\bibnamefont {Yamada}}, \ and\ \bibinfo {author}
  {\bibfnamefont {J.}~\bibnamefont {Inoue}},\ }\href {\doibase
  10.1103/physrevb.77.165117} {\bibfield  {journal} {\bibinfo  {journal}
  {Physical Review B}\ }\textbf {\bibinfo {volume} {77}} (\bibinfo {year}
  {2008}),\ 10.1103/physrevb.77.165117}\BibitemShut {NoStop}%
\bibitem [{\citenamefont {Tokatly}(2010)}]{Tokatly2010}%
  \BibitemOpen
  \bibfield  {author} {\bibinfo {author} {\bibfnamefont {I.~V.}\ \bibnamefont
  {Tokatly}},\ }\href {\doibase 10.1103/physrevb.82.161404} {\bibfield
  {journal} {\bibinfo  {journal} {Physical Review B}\ }\textbf {\bibinfo
  {volume} {82}},\ \bibinfo {pages} {161404} (\bibinfo {year} {2010})},\
  \Eprint {http://arxiv.org/abs/1004.0624} {1004.0624} \BibitemShut {NoStop}%
\bibitem [{\citenamefont {Go}\ \emph {et~al.}(2018)\citenamefont {Go},
  \citenamefont {Jo}, \citenamefont {Kim},\ and\ \citenamefont {Lee}}]{Go2018}%
  \BibitemOpen
  \bibfield  {author} {\bibinfo {author} {\bibfnamefont {D.}~\bibnamefont
  {Go}}, \bibinfo {author} {\bibfnamefont {D.}~\bibnamefont {Jo}}, \bibinfo
  {author} {\bibfnamefont {C.}~\bibnamefont {Kim}}, \ and\ \bibinfo {author}
  {\bibfnamefont {H.-W.}\ \bibnamefont {Lee}},\ }\href {\doibase
  10.1103/physrevlett.121.086602} {\bibfield  {journal} {\bibinfo  {journal}
  {Physical Review Letters}\ }\textbf {\bibinfo {volume} {121}},\ \bibinfo
  {pages} {086602} (\bibinfo {year} {2018})}\BibitemShut {NoStop}%
\bibitem [{\citenamefont {Jo}\ \emph {et~al.}(2018)\citenamefont {Jo},
  \citenamefont {Go},\ and\ \citenamefont {Lee}}]{Jo2018}%
  \BibitemOpen
  \bibfield  {author} {\bibinfo {author} {\bibfnamefont {D.}~\bibnamefont
  {Jo}}, \bibinfo {author} {\bibfnamefont {D.}~\bibnamefont {Go}}, \ and\
  \bibinfo {author} {\bibfnamefont {H.-W.}\ \bibnamefont {Lee}},\ }\href
  {\doibase 10.1103/PhysRevB.98.214405} {\bibfield  {journal} {\bibinfo
  {journal} {Phys. Rev. B}\ }\textbf {\bibinfo {volume} {98}},\ \bibinfo
  {pages} {214405} (\bibinfo {year} {2018})}\BibitemShut {NoStop}%
\bibitem [{\citenamefont {Phong}\ \emph {et~al.}(2019)\citenamefont {Phong},
  \citenamefont {Addison}, \citenamefont {Ahn}, \citenamefont {Min},
  \citenamefont {Agarwal},\ and\ \citenamefont {Mele}}]{Phong2019}%
  \BibitemOpen
  \bibfield  {author} {\bibinfo {author} {\bibfnamefont {V.~T.}\ \bibnamefont
  {Phong}}, \bibinfo {author} {\bibfnamefont {Z.}~\bibnamefont {Addison}},
  \bibinfo {author} {\bibfnamefont {S.}~\bibnamefont {Ahn}}, \bibinfo {author}
  {\bibfnamefont {H.}~\bibnamefont {Min}}, \bibinfo {author} {\bibfnamefont
  {R.}~\bibnamefont {Agarwal}}, \ and\ \bibinfo {author} {\bibfnamefont
  {E.~J.}\ \bibnamefont {Mele}},\ }\href {\doibase
  10.1103/PhysRevLett.123.236403} {\bibfield  {journal} {\bibinfo  {journal}
  {Phys. Rev. Lett.}\ }\textbf {\bibinfo {volume} {123}},\ \bibinfo {pages}
  {236403} (\bibinfo {year} {2019})}\BibitemShut {NoStop}%
\bibitem [{\citenamefont {Canonico}\ \emph {et~al.}(2020)\citenamefont
  {Canonico}, \citenamefont {Cysne}, \citenamefont {Molina-Sanchez},
  \citenamefont {Muniz},\ and\ \citenamefont
  {Rappoport}}]{canonico2020orbital}%
  \BibitemOpen
  \bibfield  {author} {\bibinfo {author} {\bibfnamefont {L.~M.}\ \bibnamefont
  {Canonico}}, \bibinfo {author} {\bibfnamefont {T.~P.}\ \bibnamefont {Cysne}},
  \bibinfo {author} {\bibfnamefont {A.}~\bibnamefont {Molina-Sanchez}},
  \bibinfo {author} {\bibfnamefont {R.}~\bibnamefont {Muniz}}, \ and\ \bibinfo
  {author} {\bibfnamefont {T.~G.}\ \bibnamefont {Rappoport}},\ }\href@noop {}
  {\bibfield  {journal} {\bibinfo  {journal} {arXiv preprint arXiv:2001.03592}\
  } (\bibinfo {year} {2020})}\BibitemShut {NoStop}%
\bibitem [{\citenamefont {Bhowal}\ and\ \citenamefont
  {Satpathy}(2020)}]{Bhowal2020}%
  \BibitemOpen
  \bibfield  {author} {\bibinfo {author} {\bibfnamefont {S.}~\bibnamefont
  {Bhowal}}\ and\ \bibinfo {author} {\bibfnamefont {S.}~\bibnamefont
  {Satpathy}},\ }\href {\doibase 10.1103/PhysRevB.101.121112} {\bibfield
  {journal} {\bibinfo  {journal} {Phys. Rev. B}\ }\textbf {\bibinfo {volume}
  {101}},\ \bibinfo {pages} {121112} (\bibinfo {year} {2020})}\BibitemShut
  {NoStop}%
\bibitem [{\citenamefont {Go}\ and\ \citenamefont {Lee}(2020)}]{go2020orbital}%
  \BibitemOpen
  \bibfield  {author} {\bibinfo {author} {\bibfnamefont {D.}~\bibnamefont
  {Go}}\ and\ \bibinfo {author} {\bibfnamefont {H.-W.}\ \bibnamefont {Lee}},\
  }\href@noop {} {\bibfield  {journal} {\bibinfo  {journal} {Physical Review
  Research}\ }\textbf {\bibinfo {volume} {2}},\ \bibinfo {pages} {013177}
  (\bibinfo {year} {2020})}\BibitemShut {NoStop}%
\bibitem [{\citenamefont {Liu}\ \emph {et~al.}(2020)\citenamefont {Liu},
  \citenamefont {Xiao}, \citenamefont {Koo},\ and\ \citenamefont
  {Yan}}]{Liu2020chiral}%
  \BibitemOpen
  \bibfield  {author} {\bibinfo {author} {\bibfnamefont {Y.}~\bibnamefont
  {Liu}}, \bibinfo {author} {\bibfnamefont {J.}~\bibnamefont {Xiao}}, \bibinfo
  {author} {\bibfnamefont {J.}~\bibnamefont {Koo}}, \ and\ \bibinfo {author}
  {\bibfnamefont {B.}~\bibnamefont {Yan}},\ }\href@noop {} {\bibfield
  {journal} {\bibinfo  {journal} {arXiv:2008.08881}\ } (\bibinfo {year}
  {2020})}\BibitemShut {NoStop}%
\bibitem [{\citenamefont {Saitoh}\ \emph {et~al.}(2006)\citenamefont {Saitoh},
  \citenamefont {Ueda}, \citenamefont {Miyajima},\ and\ \citenamefont
  {Tatara}}]{saitoh2006conversion}%
  \BibitemOpen
  \bibfield  {author} {\bibinfo {author} {\bibfnamefont {E.}~\bibnamefont
  {Saitoh}}, \bibinfo {author} {\bibfnamefont {M.}~\bibnamefont {Ueda}},
  \bibinfo {author} {\bibfnamefont {H.}~\bibnamefont {Miyajima}}, \ and\
  \bibinfo {author} {\bibfnamefont {G.}~\bibnamefont {Tatara}},\ }\href@noop {}
  {\bibfield  {journal} {\bibinfo  {journal} {Applied physics letters}\
  }\textbf {\bibinfo {volume} {88}},\ \bibinfo {pages} {182509} (\bibinfo
  {year} {2006})}\BibitemShut {NoStop}%
\bibitem [{\citenamefont {Valenzuela}\ and\ \citenamefont
  {Tinkham}(2006)}]{valenzuela2006direct}%
  \BibitemOpen
  \bibfield  {author} {\bibinfo {author} {\bibfnamefont {S.~O.}\ \bibnamefont
  {Valenzuela}}\ and\ \bibinfo {author} {\bibfnamefont {M.}~\bibnamefont
  {Tinkham}},\ }\href@noop {} {\bibfield  {journal} {\bibinfo  {journal}
  {Nature}\ }\textbf {\bibinfo {volume} {442}},\ \bibinfo {pages} {176}
  (\bibinfo {year} {2006})}\BibitemShut {NoStop}%
\bibitem [{\citenamefont {Zhao}\ \emph {et~al.}(2006)\citenamefont {Zhao},
  \citenamefont {Loren}, \citenamefont {van Driel},\ and\ \citenamefont
  {Smirl}}]{Zhao2006}%
  \BibitemOpen
  \bibfield  {author} {\bibinfo {author} {\bibfnamefont {H.}~\bibnamefont
  {Zhao}}, \bibinfo {author} {\bibfnamefont {E.~J.}\ \bibnamefont {Loren}},
  \bibinfo {author} {\bibfnamefont {H.~M.}\ \bibnamefont {van Driel}}, \ and\
  \bibinfo {author} {\bibfnamefont {A.~L.}\ \bibnamefont {Smirl}},\ }\href
  {\doibase 10.1103/PhysRevLett.96.246601} {\bibfield  {journal} {\bibinfo
  {journal} {Phys. Rev. Lett.}\ }\textbf {\bibinfo {volume} {96}},\ \bibinfo
  {pages} {246601} (\bibinfo {year} {2006})}\BibitemShut {NoStop}%
\bibitem [{\citenamefont {Huang}\ \emph {et~al.}(2012)\citenamefont {Huang},
  \citenamefont {Fan}, \citenamefont {Qu}, \citenamefont {Chen}, \citenamefont
  {Wang}, \citenamefont {Wu}, \citenamefont {Chen}, \citenamefont {Xiao},\ and\
  \citenamefont {Chien}}]{Huang2012}%
  \BibitemOpen
  \bibfield  {author} {\bibinfo {author} {\bibfnamefont {S.~Y.}\ \bibnamefont
  {Huang}}, \bibinfo {author} {\bibfnamefont {X.}~\bibnamefont {Fan}}, \bibinfo
  {author} {\bibfnamefont {D.}~\bibnamefont {Qu}}, \bibinfo {author}
  {\bibfnamefont {Y.~P.}\ \bibnamefont {Chen}}, \bibinfo {author}
  {\bibfnamefont {W.~G.}\ \bibnamefont {Wang}}, \bibinfo {author}
  {\bibfnamefont {J.}~\bibnamefont {Wu}}, \bibinfo {author} {\bibfnamefont
  {T.~Y.}\ \bibnamefont {Chen}}, \bibinfo {author} {\bibfnamefont {J.~Q.}\
  \bibnamefont {Xiao}}, \ and\ \bibinfo {author} {\bibfnamefont {C.~L.}\
  \bibnamefont {Chien}},\ }\href {\doibase 10.1103/PhysRevLett.109.107204}
  {\bibfield  {journal} {\bibinfo  {journal} {Phys. Rev. Lett.}\ }\textbf
  {\bibinfo {volume} {109}},\ \bibinfo {pages} {107204} (\bibinfo {year}
  {2012})}\BibitemShut {NoStop}%
\bibitem [{\citenamefont {Weiler}\ \emph {et~al.}(2012)\citenamefont {Weiler},
  \citenamefont {Althammer}, \citenamefont {Czeschka}, \citenamefont {Huebl},
  \citenamefont {Wagner}, \citenamefont {Opel}, \citenamefont {Imort},
  \citenamefont {Reiss}, \citenamefont {Thomas}, \citenamefont {Gross},\ and\
  \citenamefont {Goennenwein}}]{Weiler2012}%
  \BibitemOpen
  \bibfield  {author} {\bibinfo {author} {\bibfnamefont {M.}~\bibnamefont
  {Weiler}}, \bibinfo {author} {\bibfnamefont {M.}~\bibnamefont {Althammer}},
  \bibinfo {author} {\bibfnamefont {F.~D.}\ \bibnamefont {Czeschka}}, \bibinfo
  {author} {\bibfnamefont {H.}~\bibnamefont {Huebl}}, \bibinfo {author}
  {\bibfnamefont {M.~S.}\ \bibnamefont {Wagner}}, \bibinfo {author}
  {\bibfnamefont {M.}~\bibnamefont {Opel}}, \bibinfo {author} {\bibfnamefont
  {I.-M.}\ \bibnamefont {Imort}}, \bibinfo {author} {\bibfnamefont
  {G.}~\bibnamefont {Reiss}}, \bibinfo {author} {\bibfnamefont
  {A.}~\bibnamefont {Thomas}}, \bibinfo {author} {\bibfnamefont
  {R.}~\bibnamefont {Gross}}, \ and\ \bibinfo {author} {\bibfnamefont
  {S.~T.~B.}\ \bibnamefont {Goennenwein}},\ }\href {\doibase
  10.1103/PhysRevLett.108.106602} {\bibfield  {journal} {\bibinfo  {journal}
  {Phys. Rev. Lett.}\ }\textbf {\bibinfo {volume} {108}},\ \bibinfo {pages}
  {106602} (\bibinfo {year} {2012})}\BibitemShut {NoStop}%
\bibitem [{\citenamefont {Xiao}\ \emph {et~al.}(2010)\citenamefont {Xiao},
  \citenamefont {Chang},\ and\ \citenamefont {Niu}}]{Xiao2010}%
  \BibitemOpen
  \bibfield  {author} {\bibinfo {author} {\bibfnamefont {D.}~\bibnamefont
  {Xiao}}, \bibinfo {author} {\bibfnamefont {M.-C.}\ \bibnamefont {Chang}}, \
  and\ \bibinfo {author} {\bibfnamefont {Q.}~\bibnamefont {Niu}},\ }\href
  {\doibase 10.1103/revmodphys.82.1959} {\bibfield  {journal} {\bibinfo
  {journal} {Rev. Mod. Phys.}\ }\textbf {\bibinfo {volume} {82}},\ \bibinfo
  {pages} {1959 } (\bibinfo {year} {2010})}\BibitemShut {NoStop}%
\bibitem [{\citenamefont {Nagaosa}\ \emph {et~al.}(2010)\citenamefont
  {Nagaosa}, \citenamefont {Sinova}, \citenamefont {Onoda}, \citenamefont
  {MacDonald},\ and\ \citenamefont {Ong}}]{Nagaosa2010}%
  \BibitemOpen
  \bibfield  {author} {\bibinfo {author} {\bibfnamefont {N.}~\bibnamefont
  {Nagaosa}}, \bibinfo {author} {\bibfnamefont {J.}~\bibnamefont {Sinova}},
  \bibinfo {author} {\bibfnamefont {S.}~\bibnamefont {Onoda}}, \bibinfo
  {author} {\bibfnamefont {A.~H.}\ \bibnamefont {MacDonald}}, \ and\ \bibinfo
  {author} {\bibfnamefont {N.~P.}\ \bibnamefont {Ong}},\ }\href {\doibase
  10.1103/revmodphys.82.1539} {\bibfield  {journal} {\bibinfo  {journal}
  {Reviews of Modern Physics}\ }\textbf {\bibinfo {volume} {82}},\ \bibinfo
  {pages} {1539 } (\bibinfo {year} {2010})}\BibitemShut {NoStop}%
\bibitem [{\citenamefont {B\"uttiker}(1986)}]{Buttiker1986}%
  \BibitemOpen
  \bibfield  {author} {\bibinfo {author} {\bibfnamefont {M.}~\bibnamefont
  {B\"uttiker}},\ }\href {\doibase 10.1103/PhysRevLett.57.1761} {\bibfield
  {journal} {\bibinfo  {journal} {Phys. Rev. Lett.}\ }\textbf {\bibinfo
  {volume} {57}},\ \bibinfo {pages} {1761} (\bibinfo {year}
  {1986})}\BibitemShut {NoStop}%
\bibitem [{\citenamefont {Groth}\ \emph {et~al.}(2014)\citenamefont {Groth},
  \citenamefont {Wimmer}, \citenamefont {Akhmerov},\ and\ \citenamefont
  {Waintal}}]{Groth2014kwant}%
  \BibitemOpen
  \bibfield  {author} {\bibinfo {author} {\bibfnamefont {C.~W.}\ \bibnamefont
  {Groth}}, \bibinfo {author} {\bibfnamefont {M.}~\bibnamefont {Wimmer}},
  \bibinfo {author} {\bibfnamefont {A.~R.}\ \bibnamefont {Akhmerov}}, \ and\
  \bibinfo {author} {\bibfnamefont {X.}~\bibnamefont {Waintal}},\ }\href@noop
  {} {\bibfield  {journal} {\bibinfo  {journal} {New Journal of Physics}\
  }\textbf {\bibinfo {volume} {16}},\ \bibinfo {pages} {063065} (\bibinfo
  {year} {2014})}\BibitemShut {NoStop}%
\bibitem [{\citenamefont {B\"uttiker}(1988)}]{Buttiker1988}%
  \BibitemOpen
  \bibfield  {author} {\bibinfo {author} {\bibfnamefont {M.}~\bibnamefont
  {B\"uttiker}},\ }\href {\doibase 10.1147/rd.323.0317} {\bibfield  {journal}
  {\bibinfo  {journal} {IBM Journal of Research and Development}\ }\textbf
  {\bibinfo {volume} {32}},\ \bibinfo {pages} {317} (\bibinfo {year}
  {1988})}\BibitemShut {NoStop}%
\bibitem [{\citenamefont {Büttiker}(1986)}]{Buttiker1986b}%
  \BibitemOpen
  \bibfield  {author} {\bibinfo {author} {\bibfnamefont {M.}~\bibnamefont
  {Büttiker}},\ }\href {\doibase 10.1103/physrevb.33.3020} {\bibfield
  {journal} {\bibinfo  {journal} {Physical Review B}\ }\textbf {\bibinfo
  {volume} {33}},\ \bibinfo {pages} {3020} (\bibinfo {year}
  {1986})}\BibitemShut {NoStop}%
\bibitem [{\citenamefont {Papaconstantopoulos}\ \emph
  {et~al.}(1986)\citenamefont {Papaconstantopoulos} \emph
  {et~al.}}]{handbook1986}%
  \BibitemOpen
  \bibfield  {author} {\bibinfo {author} {\bibfnamefont {D.~A.}\ \bibnamefont
  {Papaconstantopoulos}} \emph {et~al.},\ }\href@noop {} {\emph {\bibinfo
  {title} {Handbook of the band structure of elemental solids}}}\ (\bibinfo
  {publisher} {Springer},\ \bibinfo {year} {1986})\BibitemShut {NoStop}%
\bibitem [{\citenamefont {Avci}\ \emph {et~al.}(2015)\citenamefont {Avci},
  \citenamefont {Garello}, \citenamefont {Ghosh}, \citenamefont {Gabureac},
  \citenamefont {Alvarado},\ and\ \citenamefont {Gambardella}}]{Avci2015}%
  \BibitemOpen
  \bibfield  {author} {\bibinfo {author} {\bibfnamefont {C.~O.}\ \bibnamefont
  {Avci}}, \bibinfo {author} {\bibfnamefont {K.}~\bibnamefont {Garello}},
  \bibinfo {author} {\bibfnamefont {A.}~\bibnamefont {Ghosh}}, \bibinfo
  {author} {\bibfnamefont {M.}~\bibnamefont {Gabureac}}, \bibinfo {author}
  {\bibfnamefont {S.~F.}\ \bibnamefont {Alvarado}}, \ and\ \bibinfo {author}
  {\bibfnamefont {P.}~\bibnamefont {Gambardella}},\ }\href {\doibase
  10.1038/nphys3356} {\bibfield  {journal} {\bibinfo  {journal} {Nature
  Physics}\ }\textbf {\bibinfo {volume} {11}},\ \bibinfo {pages} {570}
  (\bibinfo {year} {2015})},\ \Eprint {http://arxiv.org/abs/1502.06898}
  {1502.06898} \BibitemShut {NoStop}%
\end{thebibliography}
%

\end{document}